\documentstyle[aps,preprint,epsf,graphicx]{revtex}
\tighten
\draft

\begin{document}

\preprint{
\noindent
\hfill
\begin{minipage}[t]{3in}
\begin{flushright}
KRL-MAP-284\\
\vspace*{.7in}
\end{flushright}
\end{minipage}
}

\title{
The Off-diagonal Goldberger-Treiman Relation and Its Discrepancy }
\author{Shi-Lin Zhu$^{a}$, M. J. Ramsey-Musolf$^{a,b}$\\
$^a$California Institute of Technology, Pasadena, CA 91125\\
$^b$ Department of Physics, University of Connecticut,
Storrs, CT 06269 \\
}

\maketitle

\begin{abstract}

We study the off-diagonal Goldberger-Treiman relation (ODGTR) and its
discrepancy (ODGTD)
in the $N$, $\Delta$, $\pi$ sector through ${\cal O}(p^2)$
using heavy baryon chiral perturbation
theory. To this order,
the ODGTD  and axial vector $N$ to $\Delta$ transition radius are
determined solely by low energy constants. Loop
corrections appear at ${\cal O}(p^4)$. For low-energy constants of natural
size, the ODGTD
would represent a $\sim 2\%$ correction to the ODGTR. We discuss the
implications of
the ODGTR and ODGTD for lattice and quark model calculations of the
transition form factors
and for parity-violating electroexcitation of the $\Delta$.

\medskip
PACS Indices: 11.30.Rd, 13.75.Gx

\end{abstract}
\vspace{0.3cm}

\pagenumbering{arabic}

\newpage

%%%%%%%%%%%%%%%%%%%%%%%%%%%%%%%%%%%%%%%%%%%%%%%%%%%%%%%%%%%%%%%%%%%%%%%%%%%%%%%%
\section{Introduction}
\label{sec1}
%%%%%%%%%%%%%%%%%%%%%%%%%%%%%%%%%%%%%%%%%%%%%%%%%%%%%%%%%%%%%%%%%%%%%%%%%%%%%%%%

The Goldberger-Treiman relation (GTR) \cite{gt} plays an important role in
theoretical
hadronic and nuclear physics. It relates hadronic matrix elements of the
weak axial
current (the nucleon axial charge, $g_A$, and the pion
decay constant, $F_\pi$) to quantities governed by the strong interaction
(the pion nucleon
strong coupling constant,
$g_{\pi NN}$, and nucleon mass, $m_N$):
\begin{equation}\label{gold}
g_{\pi NN}={g_A m_N\over F_\pi}
\end{equation}
The GTR represents an approximation, since $g_{\pi NN}$ is determined
experimentally
at the point $q^2=m_\pi^2$ while  $g_A$ is measured close to the point
$q^2=0$.
In the chiral limit, the GTR would be exact, while in the physical world,
it holds to
an astonishing level of accuracy. The small difference between the physical
value of $g_{\pi NN}$
and RHS of Eq. (\ref{gold}) when physical values of $g_A$, $F_\pi$ and
$m_N$ are used is called the
Goldberger-Treiman discrepancy (GTD). Physically, the GTD is driven by the
explicit
chiral symmetry breaking introduced by the non-zero current quark mass.

Many
theoretical  discussions of this chiral symmetry-breaking effect have
appeared in the literature
\cite{gtd,dom,prd,barry,npa}. Recently the GTD in the context of
SU(3$)_L\times$SU(3$)_R$
chiral symmetry has been
analyzed by Goity {\em et al.}
\cite{goity} within the framework of heavy baryon chiral perturbation
theory (HB$\chi$PT)
\cite{j1,ijmpe}. These authors found that  chiral loop corrections appear
at ${\cal O}(p^4)$.
The  dominant contribution comes from the low energy counterterm
appearing in the ${\cal O}(p^3)$ Lagrangian. Their result is consistent
with more conventional
approaches where  current quark mass plays an explicit role \cite{gtd,dom}.

In this work we  analyze the off-diagonal Goldberger-Treiman relation
(ODGTR) and its
discrepancy for the $SU(2)$ $\pi$, $N$, $\Delta$ sector.  As we show below,
both the
magnitude of, and theoretical uncertainty in, the off-diagonal
Goldberger-Treiman
discrepancy (ODGTD) is $\sim m_\pi^2/\Lambda_\chi^2\sim 0.01$, where
$\Lambda_\chi=4\pi
F_\pi\sim 1$ GeV is the scale of chiral symmetry breaking. Consequently,
the ODGTR
provides a useful benchmark for both experimental and theoretical studies
of the
axial vector $N\to\Delta$ transition form factors. In principle, the
ODGTR can be tested
using charged current reactions, such as neutrino excitation of the
$\Delta$, or weak
neutral current processes, such as parity-violating (PV) electroexcitation.
These processes
are sensitive to axial vector transition  form
factors, which can be related
to the strong $\pi N\Delta$ coupling via the ODGTR. The values for these
form factors
obtained from
charged current scattering are fairly uncertain. A measurement of the PV
asymmetry for neutral
current electroexcitation will be performed at the Jefferson Lab by the G0
Collaboration\cite{jlab} in hopes of providing a more precise determination
of the axial
transition form factors.  The ODGTR  also provides a check on  lattice
QCD and hadron model
calculations of the axial transition form factors.  From either
perspective, the
theoretical analysis of the ODGTR using
HB$\chi$PT appears to be a timely endeavor.

%%%%%%%%%%%%%%%%%%%%%%%%%%%%%%%%%%%%%%%%%%%%%%%%%%%%%%%%%%%%%%%%%%%%%%%%%%%%%%%%
\section{Notations}
\label{sec2}
%%%%%%%%%%%%%%%%%%%%%%%%%%%%%%%%%%%%%%%%%%%%%%%%%%%%%%%%%%%%%%%%%%%%%%%%%%%%%%%%
We follow the standard HB$\chi$PT formalism
\cite{j1,ijmpe}
and introduce the following notations:
\begin{equation}
\Sigma =\xi^2\  ,\ \  \xi =e^{i\pi\over F_\pi}\  ,\ \  \pi ={1\over 2}
\pi^a \tau^a
\end{equation}
with $F_\pi =92.4$ MeV being the pion decay constant.
The chiral vector and axial vector currents are given by
\begin{eqnarray}\nonumber
{\cal D}_\mu & = & D_\mu +V_\mu \\ \nonumber
A_\mu &=& {i\over 2}\xi^\dag (D_\mu\Sigma ) \xi^\dag \\ \nonumber
V_\mu &=&{1\over 2}(\xi \partial_\mu \xi^\dag +\xi^\dag \partial_\mu \xi)
-{i\over 2} \xi^\dag r_\mu \xi -{i\over 2} \xi l_\mu \xi^\dag \\ \nonumber
D_\mu \Sigma &=&\partial_\mu \Sigma -ir_\mu \Sigma +i\Sigma l_\mu \\ \nonumber
r_\mu &=&{\tilde v}_\mu+a_\mu\\ \nonumber
l_\mu &=&{\tilde v}_\mu-a_\mu\\ \nonumber
F_R^{\mu\nu}&=&\partial^\mu r^\nu -\partial^\nu r^\mu-i [r^\mu, r^\nu]\\
\nonumber
F_L^{\mu\nu}&=&\partial^\mu l^\nu -\partial^\nu l^\mu-i [l^\mu, l^\nu]\\
\nonumber
f_\pm^{\mu\nu}&=&\xi F_L^{\mu\nu}\xi^\dag \pm \xi^\dag F_R^{\mu\nu} \xi \\
\nonumber
\chi &=&2 B_0 (s+i p)\\
\chi_{\pm}&=&\xi^\dag \chi \xi^\dag \pm \xi \chi^\dag \xi\ \ \ ,
\end{eqnarray}
where  $s, p, a_\mu,  {\tilde v}_\mu$ are the scalar, pseudoscalar,
pseudovector,
and vector external sources with $p=p^i \tau^i$ and $a_\mu=a_\mu^i {\tau^i\over 2}$.

For the $\Delta$, we
use the isospurion formalism, treating the $\Delta$ field $T_\mu^i(x) $ as
a vector spinor in both
spin and  isospin space \cite{hhk} with the constraint $\tau^i T_\mu^i
(x)=0$. The components of
this field are
\begin{eqnarray}\nonumber
T^3_\mu =-\sqrt{{2\over 3}}\left( \begin{array}{l} \Delta^+\\ \Delta^0
 \end{array} \right)_\mu,\\ \nonumber
  T^+_\mu =\left( \begin{array}{l}
\Delta^{++}\\
\Delta^+/\sqrt{3}
 \end{array} \right)_\mu, \\
 T^-_\mu =-\left( \begin{array}{l}
\Delta^0/\sqrt{3}\\
\Delta^-
 \end{array} \right)_\mu .
\end{eqnarray}
The field $T^i_\mu$ also satisfies the constraints for the
ordinary Schwinger-Rarita spin-${3\over 2}$ field,
\begin{equation}
\gamma^\mu T_\mu^i=0\ \ \  {\hbox{and}}\ \ \  p^\mu T_\mu^i=0\ \ \ .
\end{equation}
We eventually convert to the heavy baryon expansion, in which case the
latter constraint
becomes $v^\mu T_\mu^i=0$ with $v_\mu$ being the heavy baryon velocity.

In order to obtain proper chiral counting for the nucleon, we
employ the conventional heavy baryon expansion, and in order to
consistently include
the $\Delta$ we follow the small scale expansion developed in
\cite{hhk}. In this approach external energy and momenta,  the $\Delta$ and nucleon mass
difference
$\delta \equiv m_\Delta -m_N$ and $1/m_N$ are all treated
as ${\cal O}(\epsilon)$ in chiral power counting.

The leading order HB$\chi$PT Lagrangian reads:
\begin{eqnarray}\label{pc}\nonumber
&{\cal L}_v^{(1)} = \bar N [iv\cdot D +2g_A S\cdot A]N
-i {\bar T}^\mu_i [iv\cdot D^{ij} -\delta^{ij} \delta +g_1 S\cdot A^{ij}]
T_\mu^j \\
&+g^0_{\pi N\Delta}[{\bar T}^\mu_i \omega_\mu^i N
+ \bar N \omega_\mu^{i\dag} T_i^\mu]
+{F_\pi^2\over 4} Tr [ D^\mu \Sigma D_\mu \Sigma^\dag +\chi \Sigma^\dag
+\chi^\dag
\Sigma ]+\cdots
\end{eqnarray}
where $S_\mu$ is the Pauli-Lubanski spin operator
and $\omega_\mu^i =Tr \left( \tau^i A_\mu\right)$.

At subleading order we collect only the $\pi N \Delta$ interaction pieces
which are relevant in the following discussion.
\begin{eqnarray}\label{pc2}
{\cal L}_v^{(1)} = {1\over \Lambda_\chi}{\bar T}^\mu_i [ i {\tilde b}_3
v^\nu \omega_{\mu\nu}^i
- {{\tilde b}_8\over m_N} \omega_{\mu\nu}^i D^\nu ] N +\mbox{h.c.}
+\cdots
\end{eqnarray}
where
\begin{eqnarray}
 \omega_{\mu\nu}^i = Tr \left( \tau^i [{\cal D}_\mu, A_\nu]\right) \; .
\end{eqnarray}

%%%%%%%%%%%%%%%%%%%%%%%%%%%%%%%%%%%%%%%%%%%%%%%%%%%%%%%%%%%%%%%%%%%%%%%%%%%%%%%%
\section{Off-diagonal Goldberger-Treiman relation and its discrepancy}
\label{sec3}
%%%%%%%%%%%%%%%%%%%%%%%%%%%%%%%%%%%%%%%%%%%%%%%%%%%%%%%%%%%%%%%%%%%%%%%%%%%%%%%%

It is convenient to introduce the $\pi N \Delta$ form factor $G_{\pi
N\Delta}$ via
the effective Lagrangian:
\begin{equation}\label{G}
{\cal L}_{\pi N\Delta }=-{G_{\pi N\Delta}\over m_N}\bar \Delta_\mu^i
\partial^\mu \pi^i
N + h.c.
\end{equation}
In terms of the couplings appearing in Eq. (\ref{pc}), one has
\begin{equation}\label{ogtr}
G_{\pi N\Delta }={g_{\pi N\Delta} m_N\over F_\pi} \; ,
\end{equation}
where $g_{\pi N\Delta}$ is the renormalized $\pi N \Delta$ coupling constant.
We also express the matrix elements of the axial current  between
$\Delta^+$ and proton
in terms of the Adler form factors
\cite{jones,adl,sch}:
\begin{eqnarray}\label{47} \nonumber
&<\Delta^+ (p')| A^3_\mu |P (p)> =\bar \Delta^{+\nu}(p')
\{C_5^A (q^2) g_{\mu\nu}+{C_6^A(q^2)\over m_N^2} q_\mu q_\nu \\
&+[{C_3^A(q^2) \over m_N}\gamma^\lambda +{C_4^A(q^2)
\over m_N^2}p'^\lambda] (q_\lambda g_{\mu\nu}-q_\nu
g_{\lambda\mu}) \} u(p)\ \ \ ,
\end{eqnarray}
where we have displayed only matrix elements of the neutral component,
$A_\mu^3=\bar q \gamma_\mu\gamma_5 {\tau_3\over 2} q$, for brevity.
Experimentally, one expects contributions from $C_5^A$ to give the dominant
effect.
For future reference, we also define the off-diagonal charge radius, $r_A^2$:
\begin{equation}
r_A^2 =6 {d\over dq^2}\ln C_5^A(q^2)|_{q^2=0}
\end{equation}

To arrive at the ODGTR, it is useful first to contract Eq. (\ref{47}) with
$q^\mu$,
yielding
\begin{eqnarray}\label{48}
<\Delta^+ (p')| \partial^\mu A^3_\mu |P (p)> =i\bar \Delta^{+\nu}(p')
[C_5^A (q^2) +{C_6^A(q^2)\over m_N^2} q^2] q_\nu  u(p)\ \ \ .
\end{eqnarray}
We compute the same matrix element from the amplitudes of Fig. 1. The pion pole
contribution (Fig. 1b) depends on $G_{\pi N\Delta}(q^2)$ and $P(q^2)$, the
coupling
of the pseudoscalar current to pions. At lowest order, one has
$P(q^2)=m_\pi^2 F_\pi$.
We parameterize the non-pole contributions (Fig. 1a) in terms of a function
$C(q^2)$.
We thus obtain
\begin{eqnarray}\label{49}
<\Delta^+ (p')| \partial^\mu A^3_\mu |P (p)> =-\sqrt{2\over 3}i\bar
\Delta^{+\nu}(p')\times
{D(q^2)\over q^2-m_\pi^2 +i\epsilon} q_\nu    u(p)
\end{eqnarray}
with
\begin{equation}
\label{eq:d}
D(q^2) ={G_{\pi N\Delta}(q^2)P_\pi(q^2)\over m_N}   +(q^2-m_\pi^2) C(q^2)\
\ \ .
\end{equation}
Equating (\ref{48}) and (\ref{49}), using Eq. (\ref{eq:d}), and taking the
limit
$q^2\to 0$, leads to
\begin{equation}\label{51}
C_5^A (0) =-\sqrt{2\over 3} [-{G_{\pi N\Delta}(0)P_\pi(0)\over m_N m_\pi^2}
+C(0)]\ \ \ .
\end{equation}

We emphasize that Eq. (\ref{51}) involves no approximation. However,
neither $G_{\pi N\Delta}(0)$
nor $P_\pi(0)$ is experimentally accessible. To the extent that these
quantities vary
gently between $q^2=m_\pi^2$ and $q^2=0$ we may replace them in Eq.
(\ref{51}) with
their values at $q^2=m_\pi^2$. Assuming pion pole dominance and neglecting
$C(0)$ would then
lead to the ODGTR. The off-diagonal Goldberger-Treiman discrepancy (ODGTD),
$\Delta_\pi$,
embodies the corrections to these approximations. Including $\Delta_\pi$ we
have the
corrected ODGTR:
\begin{equation}\label{511}
C_5^A (0) =\sqrt{2\over 3}{G_{\pi N\Delta}(m_\pi^2)P_\pi(m_\pi^2)\over m_N
m_\pi^2}
(1-\Delta_\pi)
\end{equation}
where, to leading order in light-quark masses, we have
\begin{equation}\label{52}
\Delta_\pi =m_\pi^2 {d\over d q^2}\ln D(q^2)|_{q^2=m_\pi^2}\ \ \ .
\end{equation}
An analogous expression for the diagonal GTD case was first derived in Ref.
\cite{goity}.
Indeed, our treatment here largely follows the outline of that work.

In order to obtain $\Delta_\pi$, one requires the $q^2$-dependence of both
$G_{\pi
N\Delta}(q^2)$ and $P_\pi(q^2)$ as well as the non-pole amplitude $C(0)$.
To that end,
we first observe that since $P(q^2)=m_\pi^2 F_\pi$ at lowest order,
$C_5^A(0)$ starts
off as ${\cal O}(p^0)$. The non-pole term $C(0)$ generates an ${\cal
O}(p^2)$ correction,
as we discuss in the following section. In principle, since $P(q^2)$ is
${\cal O}(p^2)$
at leading order, one might expect its
$q^2$-dependence to arise at ${\cal O}(p^4)$. However, there exist no
operators in the ${\cal
O}(p^4)$ Lagrangian ( see Ref.  \cite{gasser} ) which contribute to this
$q^2$-dependence, nor do
the corresponding loop graphs contribute at this order.

The $q^2$-dependence of $G_{\pi N\Delta}(q^2)$ requires more care. As we
show explicitly
below, loop contributions to this $q^2$-dependence arise first at ${\cal
O}(p^4)$, and thus,
for our analysis, may be neglected. However, in the nonrelativistic theory
obtained via
the heavy baryon expansion, the ${\tilde b_3}+{\tilde b_8}$ terms
contribute to the
$q^2$-dependence via the factor
\begin{equation}\label{p0}
v\cdot q = {m_\Delta^2-m_N^2-q^2\over 2 m_N}\ \ \ .
\end{equation}
Note that this term is nominally ${\cal O}(p)$ in the small scale
expansion, since
$m_\Delta^2-m_N^2/2m_N\approx\delta$. However, it contains an ${\cal
O}(p^2)$ contribution
(the $q^2$ term) as a consequence of kinematics. Since we derive
expressions below valid in
the nonrelativistic theory, we should include this contribution to $G_{\pi
N\Delta}(q^2)$.

To complete  analysis of $G_{\pi N\Delta}(q^2)$, we observe
that loop corrections renormalize the bare
$\pi N \Delta$ coupling $g_{\pi N\Delta}^0 \to g_{\pi N\Delta}$ at
${\cal O}(p^2)$. However, the $q^2$-dependence of the vertex due
to loop
corrections appear ${\cal O}(p^4)$. Since we
truncate
at  ${\cal O}(p^2)$, these corrections can be neglected, and all
we need
to do is to replace $g_{\pi N\Delta}^0$ by $g_{\pi N\Delta}$.
A similar situation holds for
the diagonal GTD, as shown in the analysis of Ref. \cite{goity}.
In our case this observation directly
leads to the conclusion that the  $\Delta_\pi$ and
$r_A^2$ are solely determined by the counterterms.

It is useful to examine the $q^2$-dependence of loop effects in some detail. To
that end, we first classify the various diagrams contributing to the ODGTR.
Diagrams (a),  (e), (g),
(i), (j) and (k) contribute to the tensor structure $g_{\mu\nu}$ while the
remaining
diagrams contribute to the structure $q_\mu q_\nu$.
The first diagram (a) in FIG. 1 is the tree level one.
The second diagram (b) is the pion pole contribution. Diagram (c) and (d)
renormalize $P_\pi (q^2)$
and their contribution is of ${\cal O}(p^4)$ as explained above.
The loops in diagrams (e) and (f) contain
no $q^2$-dependence. Diagrams (g)-(n) are similar to each other, so we take
diagram (g)
as example.
The amplitude reads
\begin{eqnarray}\label{63}\nonumber
iM_{(g)}\sim {g_A^2 g_{\pi N\Delta}\over F_\pi^2} \int {d^Dk\over (2\pi)^D}
{k_\mu S\cdot q S\cdot k\over k^2-m_\pi^2+i\epsilon}\\
\times{1\over v\cdot k +{k^2-(v\cdot
k)^2\over 2m_N}}{1\over v\cdot (k+q) +{(k+q)^2-(v\cdot (
k+q))^2\over 2m_N}}
\end{eqnarray}
where $q$ is the external momentum and
we include the leading recoil correction in the nucleon propagator.
According to HB$\chi$PT, the recoil corrections may be included
perturbatively, so we expand the baryon propagators in (\ref{63}) as follows:
\begin{eqnarray}\label{65} \nonumber
iM_{(g)}\sim \int {d^Dk\over (2\pi)^D}
{k_\mu S\cdot q S\cdot k\over k^2-m_\pi^2+i\epsilon}{1\over v\cdot k}
{1\over v\cdot k+\delta}\\
\times
[1-{k^2-(v\cdot k)^2 \over  m_N (v\cdot k)}
+{(v\cdot q)^2-2k\cdot q\over 2m_N v\cdot k} +{v\cdot q\over m_N}]+\cdots
\end{eqnarray}

The first term inside the square brackets generates a $q^2$-independent
contribution
of ${\cal O}(p^2)$. Upon integration, the terms in the integrand containing
explicit
factors of $q$ generate an additional factor of $v\cdot q/m_N$ relative to
the leading
term. According to Eq. (\ref{p0}), this factor contains a $q^2$-dependent
term which
goes as $-q^2/2m_N^2$. Thus, the $q^2$-dependence of this integral occurs at
${\cal O}(p^4)$. Similar arguments hold for the other loops in diagrams
(h)-(n).

%%%%%%%%%%%%%%%%%%%%%%%%%%%%%%%%%%%%%%%%%%%%%%%%%%%%%%%%%%%%%%%%%%%%%%%%%%%%%%%%
\section{The low energy counter terms }
\label{sec4}
%%%%%%%%%%%%%%%%%%%%%%%%%%%%%%%%%%%%%%%%%%%%%%%%%%%%%%%%%%%%%%%%%%%%%%%%%%%%%%%%

Consider first $\Delta_\pi$. We collect the ${\cal O}(p^3)$
low energy counterterms which may contribute to $\Delta_\pi$:
\begin{eqnarray}\label{66}
{\cal L}^{(3)}_{CT}&=&-{c_1\over \Lambda^2_\chi} \bar T_i^\mu [D_\mu,
\chi_-]^i N
+{c_2\over \Lambda_\chi^2}\bar T^i_\mu [D_\nu, f_-^{\mu\nu}]^i N\\
\nonumber
&&+{c_3\over \Lambda^2_\chi} \bar T_i^\mu i\gamma_5  [\chi_- A_\mu]^i N
+{\rm h.c.}+\cdots
\end{eqnarray}
where $[D_\mu, \chi_-]^i=Tr\{{\tau^i\over 2} [D_\mu, \chi_-]\}$ etc.
The ellipses denotes other ${\cal O}(p^3)$ terms which do not contribute
to $\Delta_\pi$. Detailed expressions of these terms can be found in Ref.
\cite{fettes}.
After carrying out the heavy baryon expansion,
the third term in Eq. (\ref{66}) is of ${\cal O}(p^5)$, where one power of
$p$ arises from a factor of $p/m_N$ generated by
the $i\gamma_5$ tensor structrure. Also the third term
contains two pion fields. So its contribution to $\Delta_\pi$
involves with one additional loop and is further suppressed by
$1/\Lambda_\chi^2$.
In other words, this piece can be neglected.

Since we obtained our general expression for $\Delta_\pi$ using matrix
elements of
$\partial^\mu A_\mu^3$, we may deduce its dependence on the $c_i$ by varying
${\cal L}^{(3)}_{CT}$ with respect to the pseudoscalar source, $p^i$. To that
end, we use
the chiral Ward identity of QCD
\begin{equation}\label{67}
\partial^\mu [\bar q\gamma_\mu \gamma_5 {\tau^i\over 2} q] ={\hat m}
[\bar qi \gamma_5 \tau^i q]
\end{equation}
with ${\hat m}={m_u+m_d\over 2}$. Moreover,
\begin{equation}\label{68}
\bar qi \gamma_5 \tau^i q ={\delta {\cal L}_{QCD}\over \delta p^i}
\end{equation}
From Eqs. (\ref{67},\ref{68}) and the leading-order relation
$\chi^i_- =4i B_0 p^i$ we obtain
\begin{equation}\label{69}
\partial^\mu A_\mu^i =4i {\hat m}B_0 {\delta {\cal L}_{HB\chi PT}\over
\delta\chi_-^i}\ \ \ .
\end{equation}
Equations (\ref{48},\ref{49},\ref{eq:d}) and (\ref{69}) then imply that
\begin{eqnarray}\label{70}
C_5^A (q^2) +{C_6^A(q^2)\over m_N^2} q^2&=&
-\sqrt{2\over 3}\Bigl[{m_\pi^2\over q^2-m_\pi^2}
\Bigl\{g_{\pi N\Delta}\\
\nonumber
+({\tilde b_3}+{\tilde b_8})&\Bigl[ &{m_\Delta^2-m_N^2-q^2\over 2
m_N\Lambda_\chi}\Bigr]\Bigr\}
+2c_1 {m_\pi^2\over \Lambda_\chi^2}\Bigr]
\end{eqnarray}
where we have used $2B_0 {\hat m} =m_\pi^2$. With Eq. (\ref{52}) we arrive at
the off-diagonal GTD to ${\cal O}(p^2)$:
\begin{equation}\label{71}
\Delta_\pi =\left({m_\pi\over \Lambda_\chi}\right)^2\Bigl[{2c_1\over g_{\pi
N\Delta}}
-{{\tilde b_3}+{\tilde b_8}\over 2g_{\pi N\Delta}}\left({\Lambda_\chi\over
m_N}\right)\Bigr]
\ \ \ .
\end{equation}

The ODGTD -- whose scale is of order $(m_\pi/\Lambda_\chi)^2\sim 0.01$ --
depends on
three low-energy constants:
$g_{\pi N\Delta}$,
$c_1$, and ${\tilde
b_3}+{\tilde b_8}$ (we count the latter as a single constant).
Since we have scaled out explicit factors of $1/\Lambda_\chi$ in ${\cal
L}^{(2,3)}_{CT}$,
we expect these constants to be order of order unity.
In fact determinations of $g_{\pi N\Delta}$ and ${\tilde
b_3}+{\tilde b_8}$ from $\pi N$ scattering in the resonance region
yield  \cite{fettes}
\begin{eqnarray} \nonumber
& g_{\pi N \Delta} =0.98\pm 0.05 \\ \nonumber
&{\tilde b}_3 + {\tilde b}_8  = 0.59 \pm 0.10 \ \ \  .
\end{eqnarray}

Were $c_1$ also to be of order unity, we would expect
$\Delta_\pi$ to be of order a few percent.
This magnitude for $\Delta_\pi$ is consistent with previous estimates
\cite{barry,dill}.
As in the diagonal GTR the ODGTR should hold to within a few percent
accuracy, as
a consequence of  chiral symmetry.

Consider now the leading $q^2$-dependence of $C_5^A(q^2)$.
Since loops do not contribute to the $q^2$-dependence of $C_5^A(q^2)$ at
${\cal O}(p^2)$ we need consider only the tree-level contributions
generated by ${\cal L}^{(3)}_{CT}$. They are most easily obtained by
considering the dependence of ${\cal L}^{(3)}_{CT}$ on the pseudovector
source $a_\mu^i$:
\begin{equation}\label{73}
A_\mu^i ={\delta {\cal L}_{HB\chi PT}\over \delta a_\mu^i}\ \ \ .
\end{equation}
We then arrive at
\begin{eqnarray}\label{74}
C_5^A (q^2) =\sqrt{2\over 3}\Bigl[g_{\pi N\Delta} +({\tilde b_3}+{\tilde
b_8})
\left({m_\Delta^2-m_N^2-q^2\over 2 m_N\Lambda_\chi}\right)
-2c_1 {m_\pi^2\over\Lambda_\chi^2}-c_2{q^2\over
\Lambda_\chi^2}\Bigr]
\end{eqnarray}
so that
\begin{equation}\label{75}
r_A^2 =-{6\over\Lambda_\chi^2}\Bigl[{c_2 \over g_{\pi N\Delta}}+{{\tilde
b_3}+{\tilde b_8}\over
g_{\pi N\Delta}}\left({\Lambda_\chi\over m_N}\right)\Bigr]\ \ \ ,
\end{equation}
where we have dropped higher order contributions ({\em e.g.}, corrections
of order $\delta/m_N$).
From Eq. (\ref{70}) we also conclude that
\begin{eqnarray}\label{79}\nonumber
&C_6^A(q^2) =-\sqrt{2\over 3}m_N^2 g_{\pi N\Delta}[{ 1\over
q^2-m_\pi^2} -6 r_A^2 ] +{\cal O}(q^2, m_\pi^2)\\
&=-\sqrt{2\over 3}m_N F_\pi G_{\pi N\Delta}[ {1\over
q^2-m_\pi^2} -6 r_A^2 ] +{\cal O}(q^2, m_\pi^2)
\end{eqnarray}
Note the low-$q^2$ behavior of the induced off-diagonal pseudoscalar form
factor is
completely determined (once $r_A^2$ is known), since it is expressed in
terms of the
physical and measurable parameters as can be seen from the second line in Eq.
(\ref{79}).

%%%%%%%%%%%%%%%%%%%%%%%%%%%%%%%%%%%%%%%%%%%%%%%%%%%%%%%%%%%%%%%%%%%%%%%%%%%%%%%%
\section{Implications for Experiment and Theory }
\label{sec5}
%%%%%%%%%%%%%%%%%%%%%%%%%%%%%%%%%%%%%%%%%%%%%%%%%%%%%%%%%%%%%%%%%%%%%%%%%%%%%%%%

In principle, an experimental test of the ODGTR could be carried out by
drawing upon
precise measurements of $C_5^A(0)$ and $G_{\pi N\Delta}(m_\pi^2)$.
A value for $C_5^A(0)$ has been obtained from charged current neutrino
scattering from hydrogen and deuterium \cite{barish}:
\begin{equation}
\label{eq:barish}
C_5^A(0) = \frac{1}{\sqrt{3}}(2.0\pm 0.4) \ \ \ ,
\end{equation}
where the prefactor is due to relative normalization of charged and neutral
current
amplitudes.

For the strong $\pi N \Delta$ form factor, one may rely on the analysis of
$\pi N$ scattering
given in Ref. \cite{fettes}, which gives
\begin{equation}
G_{\pi N\Delta}(m_\pi^2) = 11.6 \pm  1.3\ \ \ .
\end{equation}
Substituting this result into Eq. (\ref{511}) and dropping the correction
$\Delta_\pi$ yields the leading-order ODGTR prediction for $C_5^A(0)$:
\begin{equation}
\label{eq:leading}
C_5^A(0)_{\rm l.o.} = 0.93\pm 0.10\ \ \ .
\end{equation}
A comparison of this value with the experimental result in Eq.
(\ref{eq:barish})
leads to an experimental constraint on the ODGTD:
\begin{equation}
\Delta_\pi^{\rm exp} = -0.24 \pm 0.3 \ \ \ ,
\end{equation}
where the error is dominated by the experimental error in $C_5^A(0)$.

Alternately, one may draw upon the older analysis of the K-matrix for pion
photoproduction \cite{9,13}
in the $\Delta$ resonance region to obtain
\begin{equation}
G_{\pi N\Delta}(m_\pi^2) = 14.3 \pm 0.2\ \ \ ,
\end{equation}
which implies
\begin{equation}
\Delta_\pi^{\rm exp} = 0.01 \pm 0.2\ \ \ .
\end{equation}

In both cases, the value of $\Delta_\pi^{\rm exp}$ is consistent with zero
and, thus, in line with
our expectations that the ODGTD be of order a few percent at most. At
present, however,
the uncertainty $\Delta_\pi^{\rm exp}$ is an order of magnitude larger than
one would like in
order to test this theoretical expectation. Since this uncertainty is
dominated by the
error in $C_5^A(0)$, it would be advantageous to reduce this uncertainty
through more
precise form factor measurements.

Such measurements could also reduce the present uncertainty in $r_A^2$,
which has
been determined from charged current neutrino scattering data. An empirical
parametrization of
$C_5^A(q^2)$ obtained from this data gives \cite{exp}
\begin{equation}\label{76}
C_5^A (q^2)=C_5^A(0) { 1+ 1.21 {q^2\over 2 \mbox{GeV}^2 -q^2} \over
(1-{q^2\over
M_A^2})^2}
\end{equation}
with $M_A =1.14\to 1.28$ GeV. From this parameterization, one would deduce
\begin{equation}\label{77}
{r_A^2\over 6} =({1.21\over 2}+{2\over M_A^2}) =(1.82\to 2.14) \mbox{GeV}^{-2}
\end{equation}
Accordingly we determine
\begin{equation}\label{78}
c_2 =-(3.1\to 3.5)
\end{equation}
While the value for $c_2$ is consistent
with expectations that it be of order unity, its uncertainty is roughly 10\%.

Parity-violating (PV) electroexcitation of the $\Delta$, as approved to run
at Jefferson Lab \cite{jlab}, will provide new, precise measurements of the
axial
vector $N\to\Delta$ amplitude at a variety of $q^2$ points. At first
glance, this program
of measurements could yield a determination of both $C_5^A(0)$ and $r_A^2$.
However, the
extraction of these quantities from experiment requires resolution of two
theoretical
issues. The first involves the overall normalization of the axial vector
amplitude and, thus,
the determination of $C_5^A(0)$. The normalization -- which could be
obtained from a fit to the measured $q^2$-dependence\cite{wells02} -- is
strongly
affected by electroweak radiative
corrections, $R_A^\Delta$, as discussed in detail in Ref. \cite{zmshr}.
As emphasized in that work, these corrections are theoretically uncertain,
as a result
of nonperturbative QCD effects, and the corresponding uncertainty could be on
the order of
10-20\% relative to the tree-level amplitude. The radiative
corrections always come in tandem with
axial vector amplitude for PV electroexcitation and cannot be determined
independently ({\em
e.g.}, by proper choice of kinematics or target). Thus, they introduce an
intrinsic, theoretical
uncertainty in the extraction of
$C_5^A(0)$ from this process. Given the estimated size of the uncertainty,
it appears unlikely that PV
electroexcitation will improve upon the result in Eq. (\ref{eq:barish}).

Nevertheless, determining the normalization of the axial vector amplitude
via the Jefferson Lab measurement would be interesting from another
perspective.
Because the theoretical uncertainty
in the ODGTD is considerably smaller than both the current experimental error in
$C_5^A(0)$ as well as
the estimated theoretical uncertainty in $R_A^\Delta$, one might use the ODGTR
prediction for $C_5^A(0)$, in
tandem with the normalization of the axial vector amplitude extracted from
PV electroexcitation,
to determine $R_A^\Delta$. Recently, the study of axial vector electroweak
corrections has taken
on added interest in light of the results of the SAMPLE experiment
\cite{sample}, which imply that
the magnitude of $R_A$ for elastic, PV electron scattering may be
considerably larger than
implied by theory\cite{zphr}. Understanding these corrections could have
important implications
for the interpretation of other precision electroweak measurements, such as
neutron
$\beta$-decay\cite{mckeown}, so it would be of interest to study them in
both the elastic and
inelastic channels.

A second interpretation issue involves the $q^2$-dependence of the PV
asymmetry and,
thus, the determination
of $r_A^2$. In contrast to the situation for elastic, PV electron
scattering -- where the
PV asymmetry vanishes linearly with $q^2$ at low-$|q^2|$, the asymmetry
for PV electroexcitation
contains a $q^2$-independent term. In the framework of Ref. \cite{zmhr},
this term is
characterized by a low-energy constant $d_\Delta$.
On the scale of the expected asymmetry, the magnitude of
the $d_\Delta$ contribution could be significant, particularly at low-$|q^2|$
where one would
want to determine
$r_A^2$. In order to determine the latter reliably, one also requires
knowledge of $d_\Delta$.

The second issue could, in principle, be resolved through a measurement of
$A_\gamma$, the
asymmetry for PV photoproduction of the $\Delta$. Since $A_\gamma$ is
proportional to
$d_\Delta$, and since chiral corrections to the asymmetry are small,
its measurement could remove the $d_\Delta$-related uncertainty in PV
electroexcitation.
Thus, measurements of both $A_\gamma$ and the PV electroexcitation
asymmetry at a
variety of $q^2$ points could yield values for $r_A^2$, $d_\Delta$, and
$R_A^\Delta$.

New, precise neutrino scattering experiments would complement this program.
Since
neutrino scattering probes of the axial vector transition amplitude are
free from
the large and theoretically uncertain radiative corrections entering
PV electroexcitation,
such experiments could, in principle, provide a theoretically clean
determination of
$C_5^A(0)$.
%{\bf prospects? talk with Garvey?}].

Finally, we observe that the ODGTR could provide a theoretical
self-consistency check on lattice
QCD  and hadron model computations of the axial vector $N\to\Delta$
transition form
factors. While there exist
lattice calculations of the electromagnetic $N\to\Delta$ amplitudes, the
axial vector amplitudes
remain to be computed. The lattice electromagnetic amplitudes appear to
differ significantly from
experimental values, and it would be useful to have a corresponding
comparison in the axial
vector channel. Historically, a variety of hadron model calculations of
$C_5^A(0)$ have
been performed, with predictions generally lying in the range $0.8\to 2.0$ (see
Ref. \cite{muk98} for a compilation). Those lying near the lower end of this
range are most consistent with the ODGTR, based on the value of $G_{\pi
N\Delta}(m_\pi^2)$ from Ref. \cite{fettes}. For example, the quark model
calculation of Ref. \cite{barry} predicts $C_5^A(0)$ in terms of $g_A$, and the
nucleon and $\Delta$ masses:
\begin{equation}
C_5^A(0)_{\rm Q.M.} =\frac{1}{1.17}\frac{6}{5\sqrt{3}}\left({2
m_\Delta\over m_\Delta
+ m_N}\right)
g_A = 0.87\ \ \ .
\end{equation}
The leading order ODGTR prediction is given in Eq. (\ref{eq:leading}),
where the
uncertainty is dominated by the error
in $G_{\pi N\Delta}(m_\pi^2)$ obtained from Ref. \cite{fettes}. Thus, the
quark model
appears to be consistent with the expectations derived from chiral symmetry
and the
latest analysis of strong interaction data. Having in hand similar
agreement with future
lattice calculations would be similarly satisfying.

\section*{Acknowledgment}

This work was supported in part under U.S. Department of
Energy contract \#DE-FGO2-00ER41146 and
the National Science Foundation under award PHY00-71856.
We thank J. L. Goity, B. R. Holstein, J. Martin, and S.P. Wells for
useful discussions.

%---------------------------------------------------------------------------

\vspace{2cm}

\newpage

\begin{figure}
\begin{center}
\includegraphics[height=6in,angle=270,clip=true]{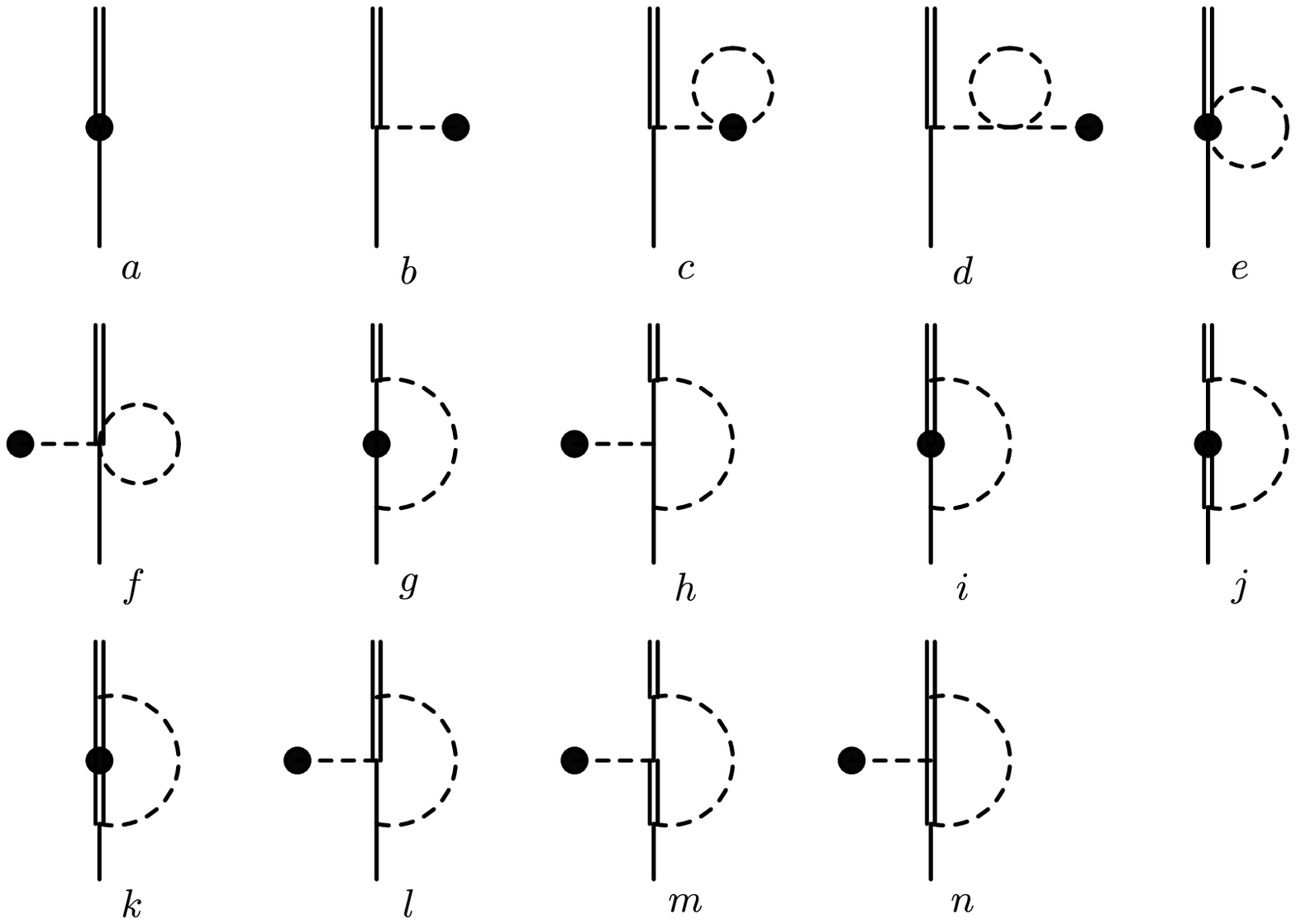}
\end{center}
\caption{\label{Fig1}
Relevant Feynman diagrams in the derivation of off-diagonal
Goldberger-Treiman relation and its discrepancy. The filled circle denotes
the pseudoscalar or pseudovector source. The double, solid and dashed lines
correspond to the delta, nucleon and pion respectively.
 }
\end{figure}

\end{document}